\documentclass{INTERSPEECH2023}
\usepackage{CJKutf8}
\usepackage{tabularx}

\interspeechcameraready

\title{Spoken Language Identification System for English-Mandarin Code-Switching Child-Directed Speech}
\name{Shashi Kant Gupta$^1$, Sushant Hiray$^2$, Prashant Kukde$^2$}
\address{
  $^1$RingCentral Innovation, India \\
  $^2$RingCentral Inc., USA
}
\email{\{shashi.gupta, sushant.hiray, prashant.kukde\}@ringcentral.com}

\begin{document}

\maketitle
 
\begin{abstract}
This work focuses on improving the Spoken Language Identification (LangId) system for a challenge that focuses on developing robust language identification systems that are reliable for non-standard, accented (Singaporean accent), spontaneous code-switched, and child-directed speech collected via Zoom. We propose a two-stage Encoder-Decoder-based E2E model. The encoder module consists of 1D depth-wise separable convolutions with Squeeze-and-Excitation (SE) layers with a global context. The decoder module uses an attentive temporal pooling mechanism to get fixed length time-independent feature representation. The total number of parameters in the model is around 22.1 M, which is relatively light compared to using some large-scale pre-trained speech models. We achieved an EER of 15.6\% in the closed track and 11.1\% in the open track (baseline system 22.1\%). We also curated additional LangId data from YouTube videos (having Singaporean speakers), which will be released for public use.
\end{abstract}

\section{Introduction}
Spoken Language Identification (LangId) plays an important role in many multi-lingual speech-related systems. Such as in home assistant devices, spoken translation, and multi-lingual speech transcription. 

Recently, with rapid progress in deep learning-based systems,  LangId has also improved significantly \cite{intro_ganapathy2014robust, intro_bartz2017language, intro_miao19b_interspeech, intro_shen2022transducer, accented_kukk2022improving, ssl_ramesh2021self}, showing SOTA performance on the VoxLingua dataset \cite{voxlingua_valk2021slt}, which is commonly used to benchmark the LangId system. Despite having significant improvements compared to early LangID systems, most of these models do not perform well on accented speech or shorter segments, making it difficult to perform well on accented code-switched speech data. Even some of the large-scale speech understanding models like wav2vec \cite{ssl_schneider2019wav2vec} and XLS-R \cite{ssl_conneau2020unsupervised}, when used for LangID in the accented scenario, do not perform well \cite{accented_kukk2022improving, ssl_ramesh2021self}. Moreover, the big size of such models makes them difficult to be used in latency-constrained or compute-constrained scenarios.

Our work focuses on improving the LangId system for the MERLIon CCS Challenge \cite{chua2023merlion} that focuses on developing robust language identification systems that are reliable for non-standard, accented (Singaporean accent), spontaneous code-switched, and child-directed speech collected via Zoom. The competition has two tracks: 1. Closed track allowing a limited set of datasets for the model training, and 2. Open track allows additional 100 hours of data and the use of pre-trained speech representation models. More details on the use of datasets are provided in \textbf{Section \ref{data_resources}}.

We proposed a very intuitive and small-size architecture for LangId (see \textbf{Section \ref{system_desc}}), which can perform well enough on smaller code-switched segments provided in the Evaluation and Development set of the competition. We compared the EER of the proposed model against some of the open-source models (see \textbf{Section \ref{results}}). Even a large-scale model like the wav2vec model gave a poor EER compared to our base model. To improve model performance on the Evaluation set, we followed several approaches. First, we curated additional training and validation split from the Development set (see \textbf{Section \ref{data_resources}}) to get a better representative training sample for the Evaluation set, which helped us improve around \textbf{~2-3\% EER} on the Evaluation set. Second, we used model ensembling for deep learning model \cite{ensembling_wang2020wisdom} to improve the proposed system to provide better prediction, which also helped improve around \textbf{~2-3\% EER} on the Evaluation set. During model ensembling, we used multiple types of loss functions, and training data splits to train the model, which helped us increase the individual model's variability (see \textbf{Sections \ref{results:closed}} and \textbf{\ref{results:open}}). We curated additional ~44 hours of training data for the open task to improve our model performance. Also, we used a pre-trained speech representations model trained on a speaker recognition task. The above two approaches helped us improve the model performance from an EER of \textbf{15.6\%} to an EER of \textbf{11.1\%}.

Summary of our major contributions are following:

\begin{itemize}
    \item We introduce a small-size architecture for LangId which shows decent performance on accented English-Mandarin Code-Switching Child-Directed Speech.
    \item We show that model pre-trained on Speaker Recognition task can significantly improve the performance on LangID.
    \item The proposed architecture can also be easily used for streaming language identification.
    \item We also show how model ensembling can improve model performance for LangId.
    \item We introduce a new Singaporean dialect English-Mandarin speech data prepared using YouTube videos. The prepared data will be released for public use on our project repo\footnote{https://github.com/shashikg/LID-Code-Switching}.
\end{itemize}

\section{Data Resources}\label{data_resources}

\subsection{Closed track}\label{data:closed}
The provided datasets for the closed track of the competition were:

\begin{itemize}
    \item 100 hours of clean speech from LibriSpeech \cite{data_panayotov2015librispeech}.
    \item 100 hours of preselected partition from the National Speech Corpus \cite{data_nsc_koh2019building}.
    \item 200 hours of preselected partition from AISHELL \cite{data_aishell_2017}.
    \item Mandarin-English Codeswitching in Southeast Asia (LDC2015S04) Corpus (SEAME Corpus) \cite{data_lyu2010seame}.
\end{itemize}

We used all the training data as it is without any additional modification, except for the National Speech Corpus dataset, which does not have speech segments of the audio files. Thus, we used an open-source Voice Activity Detection (VAD) model (marblenet \cite{jia2021marblenet} - available in Nvidia NeMo Toolkit) to generate speech segments less than or equal to 8 secs. Additionally, we created a representative training and validation split (\textbf{Val Split}) out of the provided Development Set (\textbf{Dev Set}) to improvise our model score on the Evaluation Set (\textbf{Eval Set}). We selected the split such that EER on the development set is close to the EER on the prepared validation split for a model trained without Development set data.

\subsection{Open track}\label{data:open}

For the open track, we curated two additional dataset splices. We prepared the first dataset using Mozilla Common Voice \cite{data_mcv_ardila2019common}. We only included those segments which were less than 6 secs and on which the best model from closed tasks was making errors. It gave us a total of 24.69 hours of training data having 11745 samples of Mandarin and 19315 samples of English.

\begin{CJK*}{UTF8}{gbsn}
We also collected additional speech data specific to Singaporean dialects from YouTube channels. The list of Singaporean Mandarin channels used is TiffwithMi, 老高與小茉 Mr \& Mrs Gao, iQIYI 爱奇艺, and 大聪看电影. The list of Singaporean English channels used is: JianHao Tan, Ridhwan Azman, Naomi Neo, Xiaxue, bongqiuqiu, Miss Tam Chiak, and Wah!Banana. Similar to National Speech Corpus, we used the VAD model to generate speech segments. And then, we used the model trained on the closed task to filter the dataset and use only those segments on which the model made an error. It ensured additional diversity to the training split. It gave us 19.27 hours of additional training data, with 13457 samples for Mandarin and 1543 for English. The complete list of all YouTube videos link is included in the supplementary material for reference. The curated dataset will be released for public use on our project repo\footnote{https://github.com/shashikg/LID-Code-Switching}.
\end{CJK*}

So, overall, we used an additional ~44 hours of speech data for the open track.

\section{Proposed System and Experiment Setup}\label{system_desc}
For model training, we used NeMo toolkit\footnote{https://github.com/NVIDIA/NeMo/tree/stable} \cite{kuchaiev2019nemo}. For reproducibility and to adequately describe the  model architecture and hyperparameters, we have added the model configuration for each of the cases in supplementary material. These configurations can directly be used inside the NeMo toolkit for training/finetuning purposes. We made some minor changes in the original NeMo toolkit. The changed file (label\_model.py) is also included in the supplementary material, along with the training and fine-tuning scripts. All these scripts and configurations are also available on our project repo\textsuperscript{2}.

\begin{figure}[t]
    \centering
    \includegraphics[width=\linewidth]{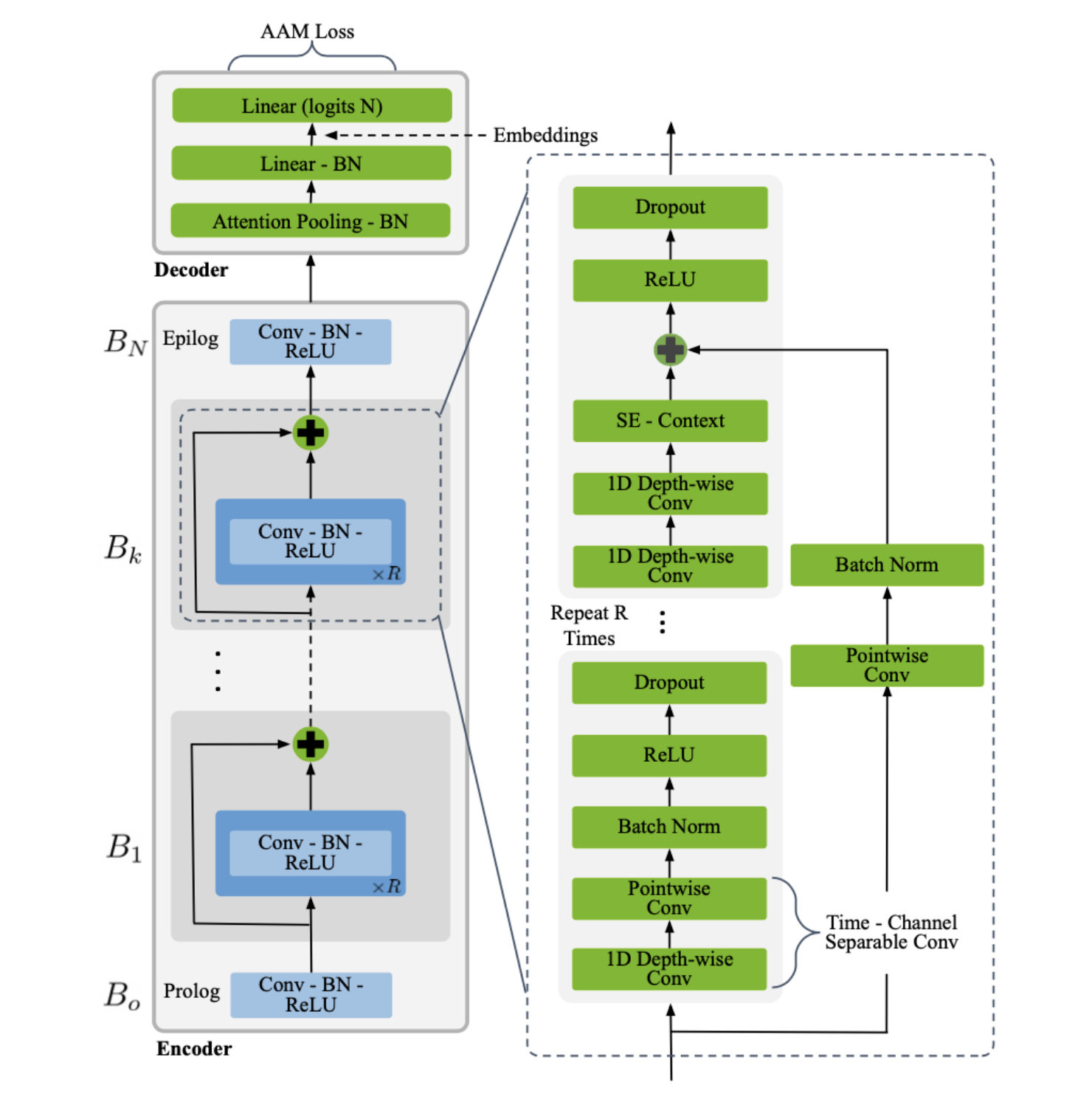}
    \caption{TitaNet Model Architecture}
    \label{fig:titanet}
\end{figure}

The model architecture is based on TitaNet architecture \cite{koluguri2022titanet} introduced for the Speaker Recognition task. We modified the last linear layer of the Decoder to produce output for only two classes corresponding to English and Mandarin language. For the closed track, we trained the model from scratch on the provided training data and further fine-tuned the model on training split curated using the Development Set (see \textbf{Section \ref{data_resources}} for details on datsets). While for the open track, we used the pre-trained model trained on the speaker recognition task, which is available in the NeMo toolkit\footnote{https://catalog.ngc.nvidia.com/orgs/nvidia/teams/nemo/\\models/titanet\_large}. For fine-tuning, we used the training split from Development Set, SEAME Corpus, Mozilla Common Voice, and YouTube scraped data as described in data resources. We also used some parts of the training data that were provided for the competition; more detail on this is provided in Section \ref{results:open}.  The encoder model consists of repeated blocks of 1D depth-wise separable convolutions with Squeeze-and-Excitation (SE) layers with a global context. The decoder is based on an attentive temporal pooling mechanism, producing a fixed length time-independent feature representation of size (batch size) X 3072. The decoded features are passed to two linear layers, one of output size 192 and another for a linear transformation from 192 to the final number of classes, i.e., 2. Instead of using attentive temporal pooling, we can also use statistical pooling layer \cite{jia2022ambernet} but we used attentive temporal pooling so that the proposed architecture can be easily adapted for streaming inference \cite{wang2022attentive}. The architecture of the TitaNet is shown in \textbf{Figure \ref{fig:titanet}} for reference purpose. Note that the small size architecture and capability to use it for streaming inference were not the limitation or criteria set for the challenge but were considered for greater impact and feasibility of the model for the real-world use case.

\subsection{Model ensembling}

In deep learning, model ensembling is not frequently used, but it poses significant advantages \cite{ensembling_wang2020wisdom}. It allows us to use smaller, lightweight models, facilitating more straightforward implementation and faster experiment iterations. It also gives us better inference speed compared to large parameter models. For ensembling, we trained multiple models with the same architecture and then collated the output probabilities from a subset of models to get the final probabilities. Individual probabilities were added together to collate the output probabilities, and then a softmax was applied to the added probabilities to get final predictions. 

To bring diversity to the prediction probabilities, we used multiple types of loss functions, and training data splits to train the model. This helped us increase the individual model’s variability and diversity in prediction probabilities. Note that out of many models trained, we picked the final subset of models for ensemble such that it performs better than any other ensemble subset on our \textit{validation split}. Details on final subset of models and individual variability cases are detailed in \textbf{Sections \ref{results:closed}} and \textbf{\ref{results:open}}.

\subsection{Preprocessing}

For the preprocessing module, acoustic features were calculated for every 25 ms frame window shifted over 10 ms. The acoustic features are 80-dimensional mel-spectrograms computed using a 512 FFT and a Hann window. Finally, the Mel-spectrogram features were normalised over the frequency axis. The sampling rate for input audio was chosen to be standard 16kHz.

\subsection{Data augmentation}

Since no additional data were allowed for the closed task, we did not use any noise or RIR to augment the training data, and instead, we used speed perturbation with a probability of 0.5 between speed rate of (0.95, 1.05) and SpecAugmentation with the following configurations: freq\_masks - 3, freq\_width - 4, time\_masks - 5, and time\_width - 0.03. 

\subsection{Loss function}\label{loss_func}

The Development and Evaluation set was highly imbalanced, so we considered testing two configurations of cross-entropy loss, one with equal weight factor for both classes and one with unequal weights. For the unequal weight case, the weight for each category was decided using the equation: \(w_x = \frac{N}{N_x}\); where \(w_x\) is weight for language \textbf{X}, \(N\) is the total number of samples, and \(N_x\) is the total number of samples of language \textbf{X}. We also tried Additive Angular Margin (AAM) \cite{deng2019arcface} loss which is supposed to help learn better intermediate features. For AAM, the loss scale was 30, and the margin was 0.01.

\subsection{Training Setup}

\subsubsection{Initial training for closed track}\label{model:init}
Initial training for the closed track on the provided training data was performed for a batch size of 32 with a limit to input audio files between 16.0 to 0.3 secs. The following parameters were used with the Adam optimizer: learning rate - 1e-3, weight decay - 1e-5, and epoch 100. We also used CosineAnnealing for LR scheduling with a warmup ratio of 0.1 and a minimum LR of 1e-7. The loss function used was cross entropy with unequal weight. To avoid picking over-fitted models, we kept the top three model checkpoints during training based on validation loss, and in the end, we picked the model with minimum validation loss.

\subsubsection{Fine-tuning for closed task}\label{model:ft_closed}
For closed tasks, the model's starting point was the output of the trained model from \textbf{Section \ref{model:init}}. Fine-tuning was performed on a training split prepared from the Development set with the following hyperparameters: batch size - 96, learning rate - 1e-4, weight decay - 1e-3, and 100 epoch. We also CosineAnnealing for LR scheduling with a warm-up ratio of 0.1 and minimum LR of 1e-6. We kept the top three model checkpoints during training based on micro accuracy on the validation set to avoid picking over-fitted models. In the final, we chose the model with minimum EER on validation split. All three types of loss functions described in \textbf{Section \ref{loss_func}} were used in the fine-tuning process. We found that cross-entropy with equal weight and AAM loss generally gives better EER, while cross-entropy with unequal weight gives better balance accuracy but poor EER. 

\subsubsection{Fine-tuning for open task}\label{model:ft_open}
For closed tasks, the model's starting point was the pre-trained model trained on speaker recognition tasks. The pre-trained checkpoints are available in the Nvidia NeMo toolkit (titanet-l\footnote{https://catalog.ngc.nvidia.com/orgs/nvidia/teams/nemo/\\models/titanet\_large} - available in Nvidia NeMo Toolkit). Fine-tuning was performed on a training split prepared from the Development set, SEAME corpus, and additional data described in \textbf{Section \ref{data:open}}. The model was trained for 100 epochs with the following hyperparameters: batch size - 96, learning rate - 1e-4, weight decay - 1e-3. We also CosineAnnealing for LR scheduling with a warm-up ratio of 0.1 and minimum LR of 1e-6. We kept the top three model checkpoints during training based on micro accuracy on the validation set to avoid picking over-fitted models. Ultimately, we chose the model with minimum EER on validation split. All three loss functions described in \textbf{Section \ref{loss_func}} were used for the fine-tuning. We found that cross-entropy with equal weight and AAM loss generally gives better EER, while cross-entropy with unequal weight gives better balance accuracy but poor EER.

\section{Results}\label{results}

We report scores (Equal Error Rate (EER) and Balanced Accuracy (BAC)) on the complete Development set (Dev Set) only for the model setup described in \textbf{Section \ref{model:init}}. While for the fine-tuned model described in \textbf{Sections \ref{model:ft_closed}} and \textbf{\ref{model:ft_open}}, we report the scores on our validation split (Val Set) that we created from the Development set and not on complete Development set because the fine-tuned model uses a part of the training split from the Development set. For results on Evaluation set (Eval Set), we have reported the result only from the best system that gave us minimum EER because of limited number of submission allowed in the challenge.

\begin{table}[th]
    \caption{EER on Closed Track}
    \label{tab:closed_task_eer}
    \centering
    \begin{tabular}{lccc}
    \toprule
    \textbf{Model}             & \textbf{Dev Set} & \textbf{Val Split} & \textbf{Eval Set} \\
    \midrule
    Base Model                 & \textbf{20.8\%}           & 20.3\%                                           & -                       \\
    SpeechBrain’s LangId       & 33.7\%                    & 33.0 \%                                           & -                       \\
    wav2vec LangId             & 24.0 \%                    & 22.4\%                                           & -                       \\
    Silero-LangId              & 42.9\%                    & 41.7\%                                           & -                       \\
    Fine-Tuned Model           & -                        & 12.7\%                                           & -                       \\
    Ensembled Fine-Tuned & -                        & \textbf{11.6\%}                                  & 15.6\%                   \\
    \bottomrule
    \end{tabular}
\end{table}

\subsection{Task 1: Closed task}\label{results:closed}

For closed task of the competition, we provide the results for three sets of models (see \textbf{Tables \ref{tab:closed_task_eer}} and \textbf{\ref{tab:closed_task_bac}}):

\begin{enumerate}
    \item \textbf{Base Model}: Model trained only on provided training data (see \textbf{Section \ref{model:init}} for more details)
    \item \textbf{Open Source Models}: For comparison, we tested the performance of some open-source models trained on a much larger corpus. The model considered were: SpeechBrain’s LangId\footnote{https://huggingface.co/speechbrain/lang-id-voxlingua107-ecapa}, a wav2vec model fine-tuned for LangId\footnote{https://huggingface.co/TalTechNLP/\\voxlingua107-xls-r-300m-wav2vec}, and Silero-LangId\footnote{https://github.com/snakers4/silero-vad/wiki/Other-Models}.
    \item \textbf{Fine-Tuned Model}: Best fine-tuned checkpoint, selected on the basis of EER on the validation split (prepared from Development Set, see \textbf{Section \ref{model:ft_closed}} for more details). The best model was trained using the cross-entropy loss with an equal weight factor.
    \item \textbf{Ensembled Fine-Tuned Model}: An ensemble of six models that gave the best score on the Validation Set. The six model constitutes the following variability cases:
    \begin{itemize}
        \item Four models were trained using cross-entropy loss with equal weight factor. Only training split from Development Set was used during fine-tuning with segments that are less than 6 secs.
        \item One other model was trained similarly to the above but also contained segments larger than 6 secs but lesser than 16 secs.
        \item One model was trained using AAM loss. It was fine-tuned using training split from Development Set as well as the SEAME corpus to increase variability. SEAME was chosen because it contains South-East Asian dialects.
    \end{itemize}
\end{enumerate}

\begin{table}[th]
    \caption{BAC on Closed Track}
    \label{tab:closed_task_bac}
    \centering
    \begin{tabular}{lccc}
    \toprule
    \textbf{Model}             & \textbf{Dev Set} & \textbf{Val Split} & \textbf{Eval Set} \\
    \midrule
    Base Model                 & 56.0\%            & 57.3\%                                           & -                       \\
    SpeechBrain’s LangId       & 67.0\%                    & 67.6\%                                           & -                       \\
    wav2vec LangId             & \textbf{67.2\%}           & 70.5\%                                           & -                       \\
    Silero-LangId              & 58.5\%                    & 54.8\%                                           & -                       \\
    Fine-Tuned Model           & -                        & 73.3\%                                           & -                       \\
    Ensembled Fine-Tuned & -                        & \textbf{75.7\%}                                  & 66.0\%                   \\
    \bottomrule
    \end{tabular}
\end{table}

\subsection{Task 2: Open task}\label{results:open}

For open task of the competition, we provide the results for two sets of models (see \textbf{Tables \ref{tab:open_task_eer}} and \textbf{\ref{tab:open_task_bac}}):

\begin{enumerate}
    \item \textbf{Fine-Tuned Model}: Best fine-tuned checkpoint, selected on the basis of EER on the validation split (prepared from Development Set, see Section \ref{model:ft_open} for more details). The best model was trained using the cross-entropy loss with an equal weight factor.
    \item \textbf{Ensembled Fine-Tuned Model}: An ensemble of seven different models which gave the best score on the Validation Set. The seven model constitutes the following variability cases:
    \begin{itemize}
        \item Four models were fine-tuned only on the training split from Development Set, YouTube scraped data, Mozilla Common Voice partition, and SEAME corpus (see \textbf{Section \ref{data:open}} for more details). Segments were less than 6 secs. Out of four, three models were trained using cross entropy with equal weight, while one was trained using AAM loss.
        \item One model was fine-tuned only on the training split from Development Set using cross entropy with equal weight.
        \item The rest of the two models were fine-tuned on all the above-mentioned datasets as well as the provided training data. Only a part of provided training data was used. They were filtered using the best model mentioned in first point. We only choose those segments on which the best model made an error.
    \end{itemize}
\end{enumerate}

\begin{table}[th]
    \caption{EER on Open Track}
    \label{tab:open_task_eer}
    \centering
    \begin{tabular}{lcc}
    \toprule
    \textbf{Model}             & \textbf{Val Split}                     & \textbf{Eval Set}           \\
    \midrule
    Fine-Tuned Model           & 9.1\%                         & -                           \\
    Ensembled Fine-Tuned Model & \textbf{7.9\%}                         & 11.1\%                      \\
    \bottomrule
    \end{tabular}
\end{table}

\begin{table}[th]
    \caption{BAC on Open Track}
    \label{tab:open_task_bac}
    \centering
    \begin{tabular}{lcc}
    \toprule
    \textbf{Model}             & \textbf{Val Split}                     & \textbf{Eval Set}           \\
    \midrule
    Fine-Tuned Model           & 82.6\%                         & -                           \\
    Ensembled Fine-Tuned Model & \textbf{83.6\%}                         & 75.7\%                      \\
    \bottomrule
    \end{tabular}
\end{table}

\section{Conclusions}
In this work, we presented a relatively small-sized architecture for LangId to improve the LangId system for a challenge that focuses on developing a robust LangId system for accented, code-switched, and child-directed speech. Apart from getting the top rank on the challenge's leaderboard, we also focused on introducing small-sized architecture and the ability for streaming inference. Our system ranked third on the closed track with an EER of 15.6\% compared to 13.9\% and 15.5\% EER of the first and second-ranked systems, respectively. On the open track also, our system ranked third with an EER of 11.1\% compared to 10.6\% and 9.5\% EER of the first and second-ranked systems, respectively. These makes the introduced system an attractive system for compute-constrained and streaming inference use cases. While having SOTA performance on accented, code-switched, and shorter audio segments.

We also curated a new speech dataset for Singaporean dialects using YouTube videos. The dataset contains English and Mandarin speech data spoken by Singaporean speakers. The curated dataset will be released for public use to drive exciting research for building robust spoken language identification for accented speech.

\bibliographystyle{IEEEtran}
\bibliography{mybib}

\begin{thebibliography}{10}
\providecommand{\url}[1]{#1}
\csname url@samestyle\endcsname
\providecommand{\newblock}{\relax}
\providecommand{\bibinfo}[2]{#2}
\providecommand{\BIBentrySTDinterwordspacing}{\spaceskip=0pt\relax}
\providecommand{\BIBentryALTinterwordstretchfactor}{4}
\providecommand{\BIBentryALTinterwordspacing}{\spaceskip=\fontdimen2\font plus
\BIBentryALTinterwordstretchfactor\fontdimen3\font minus
  \fontdimen4\font\relax}
\providecommand{\BIBforeignlanguage}[2]{{%
\expandafter\ifx\csname l@#1\endcsname\relax
\typeout{** WARNING: IEEEtran.bst: No hyphenation pattern has been}%
\typeout{** loaded for the language `#1'. Using the pattern for}%
\typeout{** the default language instead.}%
\else
\language=\csname l@#1\endcsname
\fi
#2}}
\providecommand{\BIBdecl}{\relax}
\BIBdecl

\bibitem{intro_ganapathy2014robust}
S.~Ganapathy, K.~Han, S.~Thomas, M.~Omar, M.~V. Segbroeck, and S.~S. Narayanan,
  ``Robust language identification using convolutional neural network
  features,'' in \emph{Fifteenth annual conference of the international speech
  communication association}, 2014.

\bibitem{intro_bartz2017language}
C.~Bartz, T.~Herold, H.~Yang, and C.~Meinel, ``Language identification using
  deep convolutional recurrent neural networks,'' in \emph{Neural Information
  Processing: 24th International Conference, ICONIP 2017, Guangzhou, China,
  November 14--18, 2017, Proceedings, Part VI 24}.\hskip 1em plus 0.5em minus
  0.4em\relax Springer, 2017, pp. 880--889.

\bibitem{intro_miao19b_interspeech}
X.~Miao, I.~McLoughlin, and Y.~Yan, ``{A New Time-Frequency Attention Mechanism
  for TDNN and CNN-LSTM-TDNN, with Application to Language Identification},''
  in \emph{Proc. Interspeech 2019}, 2019, pp. 4080--4084.

\bibitem{intro_shen2022transducer}
P.~Shen, X.~Lu, and H.~Kawai, ``Transducer-based language embedding for spoken
  language identification,'' \emph{arXiv preprint arXiv:2204.03888}, 2022.

\bibitem{accented_kukk2022improving}
K.~Kukk and T.~Alum{\"a}e, ``Improving language identification of accented
  speech,'' \emph{arXiv preprint arXiv:2203.16972}, 2022.

\bibitem{ssl_ramesh2021self}
G.~Ramesh, C.~S. Kumar, and K.~S.~R. Murty, ``{Self-Supervised Phonotactic
  Representations for Language Identification},'' in \emph{Proc. Interspeech
  2021}, 2021, pp. 1514--1518.

\bibitem{voxlingua_valk2021slt}
J.~Valk and T.~Alum{\"a}e, ``{VoxLingua107}: a dataset for spoken language
  recognition,'' in \emph{Proc. IEEE SLT Workshop}, 2021.

\bibitem{ssl_schneider2019wav2vec}
S.~Schneider, A.~Baevski, R.~Collobert, and M.~Auli, ``wav2vec: Unsupervised
  pre-training for speech recognition,'' \emph{arXiv preprint
  arXiv:1904.05862}, 2019.

\bibitem{ssl_conneau2020unsupervised}
A.~Conneau, A.~Baevski, R.~Collobert, A.~Mohamed, and M.~Auli, ``Unsupervised
  cross-lingual representation learning for speech recognition,'' \emph{arXiv
  preprint arXiv:2006.13979}, 2020.

\bibitem{chua2023merlion}
V.~Y.~H. Chua, H.~Liu, L.~P.~G. Perera, F.~T. Woon, J.~Wong, X.~Zhang,
  S.~Khudanpur, A.~W.~H. Khong, J.~Dauwels, and S.~J. Styles, ``Merlion ccs
  challenge: A english-mandarin code-switching child-directed speech corpus for
  language identification and diarization,'' 2023.

\bibitem{ensembling_wang2020wisdom}
X.~Wang, D.~Kondratyuk, E.~Christiansen, K.~M. Kitani, Y.~Alon, and E.~Eban,
  ``Wisdom of committees: An overlooked approach to faster and more accurate
  models,'' \emph{arXiv preprint arXiv:2012.01988}, 2020.

\bibitem{data_panayotov2015librispeech}
V.~Panayotov, G.~Chen, D.~Povey, and S.~Khudanpur, ``Librispeech: an asr corpus
  based on public domain audio books,'' in \emph{2015 IEEE international
  conference on acoustics, speech and signal processing (ICASSP)}.\hskip 1em
  plus 0.5em minus 0.4em\relax IEEE, 2015, pp. 5206--5210.

\bibitem{data_nsc_koh2019building}
J.~X. Koh, A.~Mislan, K.~Khoo, B.~Ang, W.~Ang, C.~Ng, and Y.~Tan, ``Building
  the singapore english national speech corpus,'' \emph{Malay}, vol.~20, no.
  25.0, pp. 19--3, 2019.

\bibitem{data_aishell_2017}
H.~Bu, J.~Du, X.~Na, B.~Wu, and H.~Zheng, ``Aishell-1: An open-source mandarin
  speech corpus and a speech recognition baseline,'' in \emph{2017 20th
  Conference of the Oriental Chapter of the International Coordinating
  Committee on Speech Databases and Speech I/O Systems and Assessment
  (O-COCOSDA)}, 2017, pp. 1--5.

\bibitem{data_lyu2010seame}
D.-C. Lyu, T.-P. Tan, E.~S. Chng, and H.~Li, ``Seame: a mandarin-english
  code-switching speech corpus in south-east asia,'' in \emph{Eleventh Annual
  Conference of the International Speech Communication Association}, 2010.

\bibitem{jia2021marblenet}
F.~Jia, S.~Majumdar, and B.~Ginsburg, ``Marblenet: Deep 1d time-channel
  separable convolutional neural network for voice activity detection,'' in
  \emph{ICASSP 2021-2021 IEEE International Conference on Acoustics, Speech and
  Signal Processing (ICASSP)}.\hskip 1em plus 0.5em minus 0.4em\relax IEEE,
  2021, pp. 6818--6822.

\bibitem{data_mcv_ardila2019common}
R.~Ardila, M.~Branson, K.~Davis, M.~Henretty, M.~Kohler, J.~Meyer, R.~Morais,
  L.~Saunders, F.~M. Tyers, and G.~Weber, ``Common voice: A
  massively-multilingual speech corpus,'' \emph{arXiv preprint
  arXiv:1912.06670}, 2019.

\bibitem{kuchaiev2019nemo}
O.~Kuchaiev, J.~Li, H.~Nguyen, O.~Hrinchuk, R.~Leary, B.~Ginsburg, S.~Kriman,
  S.~Beliaev, V.~Lavrukhin, J.~Cook \emph{et~al.}, ``Nemo: a toolkit for
  building ai applications using neural modules,'' \emph{arXiv preprint
  arXiv:1909.09577}, 2019.

\bibitem{koluguri2022titanet}
N.~R. Koluguri, T.~Park, and B.~Ginsburg, ``Titanet: Neural model for speaker
  representation with 1d depth-wise separable convolutions and global
  context,'' in \emph{ICASSP 2022-2022 IEEE International Conference on
  Acoustics, Speech and Signal Processing (ICASSP)}.\hskip 1em plus 0.5em minus
  0.4em\relax IEEE, 2022, pp. 8102--8106.

\bibitem{jia2022ambernet}
F.~Jia, N.~R. Koluguri, J.~Balam, and B.~Ginsburg, ``Ambernet: A compact
  end-to-end model for spoken language identification,'' \emph{arXiv preprint
  arXiv:2210.15781}, 2022.

\bibitem{wang2022attentive}
Q.~Wang, Y.~Yu, J.~Pelecanos, Y.~Huang, and I.~L. Moreno, ``Attentive temporal
  pooling for conformer-based streaming language identification in long-form
  speech,'' \emph{arXiv preprint arXiv:2202.12163}, 2022.

\bibitem{deng2019arcface}
J.~Deng, J.~Guo, N.~Xue, and S.~Zafeiriou, ``Arcface: Additive angular margin
  loss for deep face recognition,'' in \emph{Proceedings of the IEEE/CVF
  conference on computer vision and pattern recognition}, 2019, pp. 4690--4699.

\end{thebibliography}

\end{document}